\documentclass[12pt]{iopart}

\usepackage{iopams}
\usepackage{graphicx,amssymb}
\usepackage{color}

\newtheorem{proposition}{Proposition}
\newtheorem{theorem}{Theorem}

\newtheorem{definition}{Definition}
\newcommand{\be}{\begin{equation}}
\newcommand{\ee}{\end{equation}}

\newcommand{\ci}{\mathop{\textrm{i}}}

\begin{document}

\title[Radiative gravitational fields] {On the
super-energy radiative gravitational fields}

\author{Joan Josep Ferrando$^1$\
and Juan Antonio S\'aez$^2$}

\address{$^1$\ Departament d'Astronomia i Astrof\'{\i}sica, Universitat
de Val\`encia, E-46100 Burjassot, Val\`encia, Spain.}

\address{$^2$\ Departament de Matem\`atiques per a l'Economia i l'Empresa,
Universitat de Val\`encia, E-46022 Val\`encia, Spain}

\ead{joan.ferrando@uv.es; juan.a.saez@uv.es}

\begin{abstract}
We extend our recent analysis ({\it Class. Quantum Grav.} {\bf 29}
075012) on the Bel radiative gravitational fields to the
super-energy radiative gravitational fields defined by
Garc\'{\i}a-Parrado ({\it Class. Quantum Grav.} {\bf 25} 015006). We
give an intrinsic characterization of the {\it new radiative fields}, and
we consider some distinguished classes of both radiative and non
radiative fields. Several super-energy inequalities are improved.
\end{abstract}


\submitto{\CQG}

%

\section{Introduction}
\label{sec-intro}

Elsewhere \cite{fs-RS} we have analyzed the Bel concept of intrinsic
radiative gravitational field \cite{bel-1, bel-2, bel-3}, and we
have shown that the three radiative types, $N$, $III$ and $II$,
correspond with three different physical situations: pure radiation,
asymptotic pure radiation and generic (non-pure, non-asymptotic
pure) radiation. In the aforementioned paper we have also shown
that, for Bel non radiative fields, the minimum value of the
relative super-energy is acquired by the observers at rest with
respect to the field (those seeing a vanishing super-Poynting
vector).

Following Bel's ideas, Garc\'{\i}a-Parrado \cite{garcia-parrado} has
introduced new relative super-energy quantities and has written the
full set of equations for these super-energy quantities. This study
leads naturally to a concept of intrinsic radiation which is less
restrictive than that given by Bel.

Here we extend our analysis \cite{fs-RS} on the Bel approach to the
Garc\'{\i}a-Parrado radiative gravitational fields. We show that the
non radiative fields correspond to type $D$ metrics and some classes
of type $I$ metrics already considered in the literature: the $IM^+$
and the $IM^{\infty}$ metrics \cite{mcintosh}. These classes appear
in a natural way when classifying the Bel-Robinson tensor as an
endomorphism \cite{fsBR-1, fsBR-2}. Moreover, in these spacetimes
the four null Debever directions span a 3-plane \cite{mcintosh,
fs-I, fsWaligned}. On the other hand, we show that three classes of
Garc\'{\i}a-Parrado type $I$ radiative fields can be considered: the
$IM^-$, the $IM^{-6}$ and the generic type $I_r$ metrics.

Our study deepens our understanding of the Bel-Robinson tensor and
performs the Bel and Garcia-Parrado concepts. The interest of this
subject and the connection between super-energy and gravitational
energy have been widely remarked and analyzed in the literature (see
\cite{seno} and \cite{garcia-parrado}, and references therein). Now
we want to add two brief remarks.

The Bel super-energy density is the leading-order contribution to
the quasi-local energy in vacuum \cite{horowitz-schmidt}. Is there a
similar property for the other super-energy quantities? This is a
question to be analyzed in the future. The quasi-local energy is
associated with a proper energy surface density \cite{brown-york}.
Is it possible to define quasi-local quantities associated with a
proper momentum surface density or with a spatial stress? The
leading-order contribution of these (tensorial) quasi-local
quantities could be given by (tensorial) relative super-energy
quantities.

The asymptotic behavior of the field created by an isolated system
is given by the Sachs' peeling theorem \cite{sachs}. Far from the
sources or in the transitional zone, the field has a Bel radiative
behavior, type $N$ or types $III$ and $II$, respectively. However,
terms of Bel non radiative type ($I$ or $D$) are dominant near the
sources. To analyze if these terms correspond to Garcia-Parrado
radiative or non-radiative fields is a further question to be
considered. Another possible approach to this subject could be to
study a peeling-like theorem for the Bel-Robinson tensor itself. In
this approach the algebraic study of the Bel-Robinson tensor
\cite{fsBR-1, fsBR-2} will play an important role.

The paper is organized as follows. In section \ref{sec-2} we
introduce the basic concepts and notation and summarize previous
results which help us to understand the present paper. The
super-energy inequalities presented in \cite{fs-RS} are revisited in
section \ref{sec-3}, where we extend the kind of bounds already
known for the super-energy density to all the tensorial quantities
defined by contracting the Bel-Robinson tensor with the observer
velocity.
In section \ref{sec-4} we define the {\it proper super-energy
scalars} and show that, for Bel non radiative fields, they are
acquired by the observers at rest and, for Bel radiative fields, it
is the infimum for all the observers of the super-energy scalars. We
also define the principal super-stresses of a Bel non radiative
field.
Section \ref{sec-5} is devoted to studying both the
Garc\'{\i}a-Parrado radiative and non-radiative gravitational
fields. The relationships between the different  classes and both,
the Debever directions and the algebraic properties of the
Bel-Robinson tensor, are outlined.  Both, radiative and non
radiative classes are characterized in terms of the principal
super-stresses.
In section \ref{sec-6} we analyze the results of the paper with
several diagrams which clarify the relation between the different
classes of type $I$ fields, and we introduce a radiation scalar
which measures how radiative (in the sense defined by
Garcia-Parrado) is a gravitational field at a point.
Finally, we present three appendices. The first one summarizes the
algebraic classes of the Bel-Robinson tensor, the second one
presents some constraints on the relative super-energy quantities,
and the third one gives accurate proof of propositions \ref{prop-tt}
and \ref{prop-qq}. The notation that we use in this work is the same
as that used in \cite{fs-RS}.

\section{The Bel approach to radiative gravitational states}
\label{sec-2}

With the purpose of defining intrinsic states of gravitational
radiation, Bel \cite{bel-1, bel-2, bel-3} introduced the {\it
super-energy Bel tensor} which plays an analogous role for
gravitation to that played by the Maxwell-Minkowski tensor for
electromagnetism. In the vacuum case this {\it super-energy Bel
tensor} is divergence-free and it coincides with the {\it
super-energy Bel-Robinson tensor} $T$.

Using tensor $T$, Bel defined the relative {\it super-energy
density} and the {\it super-Poynting vector} associated with an
observer. Then, following the analogy with electromagnetism, the
intrinsic radiative gravitational fields are those for which the
Poynting vector does not vanish for any observer \cite{bel-1,bel-3}.

This analogy with electromagnetism also plays a fundamental role in
our analysis \cite{fs-RS} of the Bel radiative and non-radiative
fields. Now we summarize our main results introducing the basic
concepts required in the present work.

\subsection{The Bel-Robinson tensor. Algebraic restrictions}
\label{subsec-2a}

In terms of the Weyl tensor $W$, the Bel-Robinson tensor takes the
expression \cite{bel-1, bel-2, bel-3}:
\begin{equation} \label{BR-1}
{T_{\alpha  \mu \beta \nu}} = \frac14 \left({{{
W_{\alpha}}^{\rho}}_{\beta}}^{ \sigma} W_{\mu \rho \nu \sigma } +
{{{* W_{\alpha}}^{\rho}}_{\beta}}^{ \sigma} *\! W_{\mu \rho \nu
\sigma }\right) \, ,
\end{equation}

For any observer $u$, the relative {\it electric} and {\it magnetic}
Weyl fields are given by $E= W(u;u)$ and $H= *W(u;u)$, respectively.
The following relative super-energy quantities can be defined:
\begin{equation} \label{relative}
\hspace{-2.3cm} \tau = T(u,u,u,u)\,  ,\ \ \ q_{\perp} =
-T(u,u,u)_{\perp} \, ,\ \ \ t_{\perp} = T(u,u)_{\perp}\,  ,\ \ \
Q_{\perp} = -T(u)_{\perp} \, , \ \ \ T_{\perp} \, .
\end{equation}
where, for a tensor $A$, $A_{\perp}$ denotes the orthogonal
projection defined by the projector $\gamma = u \otimes u + g$.

Bel introduced the {\it super-energy density} $\tau$ and the {\it
super-Poynting (energy flux) vector} $q_{\perp}$ years ago
\cite{bel-1, bel-3}. Bonilla and Senovilla \cite{bonilla-sen} used
$t_{\perp}$ in studying the causal propagation of gravity and,
recently, Garc\'{\i}a-Parrado \cite{garcia-parrado} has considered
$Q_{\perp}$ and $T_{\perp}$. These last three relative quantities
have been called the {\it super-stress tensor}, the {\it stress flux
tensor} and the {\it stress-stress tensor}, respectively. The
expression of the relative super-energy quantities (\ref{relative})
in terms of the electric and magnetic Weyl tensors can be found in
\cite{garcia-parrado}.

In vacuum, the Bianchi identities imply that $T$ satisfies $\nabla
\cdot T=0$. For any observer, this equation shows that the relative
quantities $q_{\perp}$ and $Q_{\perp}$ play the role of fluxes of
the relative quantities $\tau$ and $t_{\perp}$, respectively
\cite{garcia-parrado}.

The algebraic constraints on the Bel-Robinson tensor playing a
similar role to that played by the Rainich conditions \cite{rainich}
for the electromagnetic energy tensor were obtained by Bergqvist and
Lankinen \cite{bergqvist-lan}.

On the other hand, we have studied elsewhere \cite{fsBR-1,fsBR-2}
the Bel-Robinson tensor $T$ as an endomorphism on the 9-dimensional
space of the traceless symmetric tensors. Its nine eigenvalues
depend on the three (complex) Weyl eigenvalues $\{ \rho_k \}$ as
$t_k = | \rho_k |^2$, $\tau_k = \rho_i \bar{\rho}_j$, $(ijk)$ being
an even permutation of $(123)$. Three independent invariant scalars
can be associated with $T$. In fact, the nine eigenvalues $\{ t_i ,
\tau_{i} , \bar\tau_i \}$ can be written in terms of three scalars
$\{ \kappa_i \}$ as \cite{fsBR-1}:
\begin{equation} \label{ti-pi}
t_i = 2 (\kappa_j + \kappa_k), \quad \tau_i = -2(\kappa_i + \ci
\kappa), \quad \kappa^2 = \kappa_1 \kappa_2 + \kappa_2 \kappa_3 +
\kappa_3 \kappa_1 \, .
\end{equation}
We have also intrinsically characterized the algebraic classes of
$T$ \cite{fsBR-1}, and we have given their Segr\`e type and their
canonical form \cite{fsBR-2}. Some of these results which we need
here are summarized in \ref{A-Bel-Robinson}. In what follows we will
make use of the scalar invariants  $\alpha$, $\xi$ and $\chi$
defined by the expressions:
\begin{eqnarray}
\hspace{-8mm} \alpha \equiv \frac12 \sqrt{(T,T)} \, ,  \quad  (T,T)
= T_{\alpha \beta \lambda \mu} T^{\alpha \beta \lambda \mu} =
\frac{1}{64}[(W,W)^2 + (W,*W)^2] \geq 0 \, ; \label{alpha}\\
\hspace{-8mm} \xi  \equiv \frac14  \sum_{i=1}^{3} t_i  \, ; \qquad
\quad  \chi  \equiv \frac14 \sum_{i=1}^{3} t_i^2  \, ; \qquad \quad
8 \xi^2 + \alpha^2 = 6 \chi \, , \label{xi-chi}
\end{eqnarray}
where the above constraint between the invariants $\alpha$, $\chi$
and $\xi$ is a consequence of the restrictions (\ref{ti-pi}) on the
Bel-Robinson eigenvalues.

\subsection{Bel radiative gravitational fields}
\label{subsec-c}

The super-energy density $\tau$ vanishes only when the Weyl tensor
$W$ vanishes. Then, if we consider $\tau$ as a measure of the
gravitational field, its flux $q_{\perp}$ denotes the presence of
gravitational radiation. This is the point of view of Bel
\cite{bel-1, bel-3}, who gave the following definition.

\begin{definition} \label{def-bel-g}
{\bf (Intrinsic gravitational radiation, Bel 1958)} In a vacuum
spacetime there exists {\em intrinsic gravitational radiation} (at a
point) if the super-Poynting vector $q_{\perp}$ does not vanish for
any observer.
\end{definition}

It is known that the Bel radiative gravitational fields are those of
Petrov-Bel types $N$, $III$ and $II$ \cite{bel-3}. Then, also
motivated by the Lichnerowicz ideas \cite{lich}, we have proposed to
distinguish three physical situations \cite{fs-RS}: the {\em pure
gravitational radiation} (type $N$), the {\em asymptotic pure
gravitational radiation} (type $III$), and the {\em generic
radiative states} (type $II$).

\subsection{Bel non radiative gravitational fields. Observer at rest and
proper super-energy density} \label{subsec-2d}

From Bel's point of view, non radiative gravitational fields are
those for which an observer exists who sees a vanishing relative
super-Poynting vector. The following definition naturally arises:

\begin{definition} \label{def-or-g}
The observers for whom the super-Poynting vector vanishes, are said
to be {\em observers at rest} with respect to the gravitational
field.
\end{definition}
It is known \cite{bel-3} that the non radiative gravitational fields
are the Petrov-Bel type $I$ or $D$ spacetimes, and the observers at
rest with respect to the gravitational field are those for whom the
electric and magnetic Weyl tensors simultaneously diagonalize. In a
type $I$ spacetime a unique observer $e_0$ at rest with respect the
gravitational field exists, and in a type $D$ spacetime the
observers $e_0$ at rest with respect to the gravitational field are
those lying on the Weyl principal plane.

In \cite{fs-RS} we have given the following definition and result:

\begin{definition} \label{def-pe-g}
We call {\em proper super-energy density} of a gravitational field
the invariant scalar $\xi$ given in (\ref{xi-chi}).

\end{definition}

\begin{theorem} \label{theo-xi}
For a Bel non radiative gravitational field (I or D) the minimum
value of the relative super-energy density is the proper
super-energy density $\xi$, which is acquired by the observers at
rest with respect to the field.

For Bel radiative gravitational field (N, III or II) the
super-energy density decreases and tends to the proper super-energy
density $\xi$ as the velocity vector of the observer approaches the
unique fundamental direction $\ell$.

For pure and asymptotic pure radiation (N or III), the proper
super-energy density $\xi$ is zero. For generic radiation (type
$II$), $\xi$ is strictly positive.
\end{theorem}

\section{Super-energy inequalities}
\label{sec-3}

The Bel-Robinson tensor satisfies the generalized dominant energy
condition \cite{seno, seno2} which implies that for any observer
$u$, the relative quantities $\tau$ and $q = - T(u,u,u)$ are subject
to the known inequalities, $\tau \geq 0$, $(q,q) \leq 0$. In
\cite{fs-RS} we have generalized these {\it super-energy
inequalities} in two aspects. On the one hand we have shown that
$\tau$ and $(q,q)$ are bounded by scalars depending on the main
quadratic invariant $\alpha$ given in (\ref{alpha}).

On the other hand we have extended these kind of bounds to other
spacetime relative quantities (see theorem 2 in \cite{fs-RS}). Now
we present stronger inequalities on these spacetime quantities,
restrictions that lead naturally to the concept of super-energy
scalars.

From the algebraic properties of the Bel-Robinson tensor and, in
particular from the Bergqvist and Lankinen conditions
\cite{bergqvist-lan}, we obtain the following propositions (see
proof in \ref{C-propositions}):

\begin{proposition} \label{prop-tt}
Let $\alpha$ and $\chi$ be the Bel-Robinson invariants defined in
(\ref{alpha}) and (\ref{xi-chi}), and $t= T(u,u)$ for any observer
$u$. Then, it holds:
\begin{itemize}
\item[-] For type $N$,  $\ \ (t,t)=\chi = \frac{1}{2} \alpha^2  =0$.
\item[-] For type $III$, $\ \ (t,t) > \chi = \frac{1}{2} \alpha^2 =0  $.
\item[-] For type $II$,  $\ \ (t,t) > \chi = \frac{1}{2} \alpha^2 > 0  $.
\item[-] For type $D$, $\ \ (t,t) \geq  \chi =  \frac{1}{2} \alpha^2 > 0  $.
\item[-] For type $I$, $\ \ (t,t) \geq  \chi \geq  \frac{1}{2} \alpha^2 \geq 0
$.
\end{itemize}
Moreover: (i) for types I and D, $\; (t_0, t_0) = \chi$, with $t_0=
T(e_0,e_0)$, $e_0$ being a principal observer, and (ii) for types
III and II, $\; (t,t)$ tends to $\chi$ as the velocity vector of the
observer approaches the unique fundamental direction $\ell$.
\end{proposition}

\begin{proposition} \label{prop-qq}
Let $\alpha$ and $\xi$ be the Bel-Robinson invariants defined in
(\ref{alpha}) and (\ref{xi-chi}), and $q= -T(u,u,u)$ for any
observer $u$. Then, it holds:
\begin{itemize}
\item[-] For type $N$,  $\ \ (q,q)= -\xi^2 = -\frac{1}{4} \alpha^2  =0$.
\item[-] For type $III$, $\ \ (q,q) < -\xi^2 = -\frac{1}{4} \alpha^2 =0  $.
\item[-] For type $II$,  $\ \ (q,q) < -\xi^2 = -\frac{1}{4} \alpha^2 < 0  $.
\item[-] For type $D$, $\ \ (q,q) \leq  -\xi^2 =  -\frac{1}{4} \alpha^2 < 0  $.
\item[-] For type $I$, $\ \ (q,q) \leq  -\xi^2 \leq  -\frac{1}{4} \alpha^2 \leq 0
$.
\end{itemize}
Moreover: (i) for types I and D, $\; (q_0, q_0) = - \xi^2$, with
$q_0= -T(e_0,e_0,e_0)$, $e_0$ being a principal observer, and (ii)
for types III and II, $\; (q,q)$ tends to $\; - \xi^2$ as the
velocity vector of the observer approaches the unique fundamental
direction $\ell$.
\end{proposition}

Now we can state the following theorem.

\begin{theorem} \label{theo-sei}
{\bf (Super-energy inequalities)} Let $T$ be the Bel-Robinson tensor
and for any observer $u$ let us define the relative spacetime
quantities:
\begin{equation} \label{T-Q-t-q-tau}
Q = -T(u)\,  ,\quad t = T(u,u) \,  ,\quad q = -T(u,u,u) \, ,\quad
 \tau = T(u,u,u,u)  \, .
\end{equation}
Then, the following super-energy inequalities hold:
\begin{equation} \label{sei}
\hspace{-1.5cm}
\begin{array}{c}
(T,T) \equiv 4 \alpha^2 \geq 0 \,  ,  \qquad (Q,Q) = - \alpha^2 \leq
0 \, , \\[3mm]  (t,t) \geq \chi  \geq \frac12 \alpha^2 \geq 0 \, ,  \ \quad
(q,q) \leq - \xi^2 \leq -\frac14 \alpha^2 \leq 0 \, ,  \ \quad  \tau
\geq \xi \geq \frac12 \alpha \geq 0 \, .
\end{array}
\end{equation}
where $\xi$ and $\chi$ are the invariant scalars defined in
(\ref{xi-chi}).
\end{theorem}

The first, second and last conditions in (\ref{sei}) have been
stated and shown in \cite{fs-RS} (theorems 1 and 2). The third
condition in (\ref{sei}) is a consequence of proposition
\ref{prop-tt} and, finally, the fourth condition in (\ref{sei}) is a
consequence of proposition \ref{prop-qq}.

\section{Lower bounds on the super-energy scalars and proper
super-energy scalars} \label{sec-4}

For any observer $u$, the amounts of the super-energy quantities
(\ref{relative}) are the {\it super-energy scalars} given by
$$
\tau  \, , \quad  |q_{\perp}| = \sqrt{(q_{\perp}, q_{\perp})} \, ,
\quad |t_{\perp}| = \sqrt{(t_{\perp}, t_{\perp})}  \, , \quad
|Q_{\perp}| = \sqrt{(Q_{\perp}, Q_{\perp})} \, , \quad   |T_{\perp}|
= \sqrt{(T_{\perp}, T_{\perp})}  \, .
$$

Note that $ \tr T_{\perp} = t_{\perp}$ and $\tr t_{\perp} = \tau$
and, consequently, the Weyl tensor vanishes when $T_{\perp}$ or
$t_{\perp}$ vanish \cite{garcia-parrado}. A similar property holds
for the super-energy scalars $|T_{\perp}|$ and $|t_{\perp}|$ because
they are the modulus of the spatial tensors $T_{\perp}$ and
$t_{\perp}$, respectively.

Theorem \ref{theo-xi} shows that the proper super-energy density
$\xi$ is the infimum,  for all the observers $u$, of the
super-energy densities $\tau_u$. This property justifies the
following definition.

\begin{definition} \label{def-proper}
We call {\em proper super-energy scalars} the
infimum  for all the observers of the super-energy scalars.
\end{definition}

The two last inequalities in (\ref{sei}) imply:
\begin{equation} \label{sec-a}
\tau^2 \geq \tau^2 - |q_{\perp}|^2 \geq \xi^2 \, , \qquad
|q_{\perp}| \geq 0 \, .
\end{equation}
On the other hand, from (\ref{sec-a}) and the quadratic scalar
constraints (\ref{qsc}) we obtain:
\begin{eqnarray}
\displaystyle |t_{\perp}|^2 = \frac13 \, \tau^2 + \frac16 \,
\alpha^2 + \frac23 \, |q_{\perp}|^2 \geq \, \frac13 \xi^2 + \frac16
\alpha^2  \, ,  \nonumber \\[1mm]
|Q_{\perp}|^2  = 2 \, \tau^2 - \frac12 \, \alpha^2  - \,
|q_{\perp}|^2 \geq \, 2 \xi^2 - \frac12 \alpha^2 \, ,  \label{sec-b}\\[1.7mm]
|T_{\perp}|^2 = 5 \, \tau^2 + \alpha^2 - 4 \, |q_{\perp}|^2 \geq \,
5 \xi^2 +  \alpha^2 \nonumber \, .
\end{eqnarray}
Moreover, as a consequence of proposition \ref{prop-qq}, equalities
in (\ref{sec-a}) and (\ref{sec-b}) hold for the observers at rest in
the case of Bel non radiative fields and, for Bel radiative fields,
the inequalities (\ref{sec-b}) approach an equality as the velocity
vector of the observer approaches the unique fundamental direction
$\ell$. Consequently, we can state the following theorem.

\begin{theorem} \label{minimum}
The proper scalars of radiated super-energy $\xi_q$, of super-stress
$\xi_t$, of radiated super-stress $\xi_Q$ and stress-stress $\xi_T$
are, respectively, the invariant scalars:
$$
\xi_q = 0 \, , \qquad  \xi_t = \sqrt{\frac13 \xi^2 + \frac16
\alpha^2} \, , \qquad  \xi_Q = \sqrt{2 \xi^2 - \frac12 \alpha^2} \,
, \qquad  \xi_T = \sqrt{5 \xi^2 +  \alpha^2}   \, .
$$
where $\xi$ and $\alpha$ are the proper energy density
(\ref{xi-chi}) and the main quadratic scalar (\ref{alpha}).

For a Bel non radiative gravitational field (I or D) the proper
super-energy scalars are acquired by the observers at rest with
respect to the field: if $e_0$ is such an observer, then $ \tau_0 =
\xi \, , \ |q_{0\perp}| = \xi_q = 0 \, , \ |t_{0\perp}| = \xi_t \, ,
\ |Q_{0\perp}| = \xi_Q  \, , \ |T_{0\perp}| = \xi_T \, $.

For a Bel radiative gravitational field (N, III or II) the scalar
associated with each relative super-energy quantity decreases and
tends to the proper scalar of this quantity as the velocity vector
of the observer approaches the unique fundamental direction $\ell$.

\end{theorem}

On the other hand, from expressions (\ref{sec-b}) we obtain a result
which extends a previous one given in \cite{bonilla-sen}:

\begin{proposition}
The super-energy scalars are subject to
the following constraints:
\begin{equation} \label{sec-c}
|Q_{\perp}|^2 - |q_{\perp}|^2 = 3(\tau^2 - |t_{\perp}|^2) \geq 0 \,
, \qquad 3 \tau - |T_{\perp}| \geq 0 \, .
\end{equation}
\end{proposition}

Note that for Bel non radiative fields (types $I$ and $D$) we can
consider the super-energy quantities relative to the observers $e_0$
at rest with respect to the field: the {\it proper super-energy}
$\tau_0 = \xi$, the {\it proper Poynting vector} $q_{0\perp} = 0$,
the {\it proper super-stress tensor} $t_{0\perp}$, the {\it proper
super-stress flux tensor} $Q_{0\perp}$ and the {\it proper
stress-stress tensor} $T_{0\perp}$. If $\{e_0, e_i\}$ is a Weyl
canonical frame of a type $I$ or type $D$ spacetime, from the
Bel-Robinson canonical form (see \cite{fs-RS}) we obtain:
\begin{eqnarray}
\displaystyle t_{0\perp} =  \sum_{i=1}^3 \kappa_i \, e_i \otimes e_i
\, , \qquad 4 \kappa_i = t_k + t_j - t_i \, , \qquad  i \not=j
\not=k\not=i \, , \label{t-0perp}  \\
\displaystyle Q_{0\perp} =  \kappa\, \sum_{\sigma}(e_{\sigma(1)}
\otimes e_{\sigma(2)} \otimes e_{\sigma(3)}) \, , \qquad \kappa^2 =
\kappa_1 \kappa_2 + \kappa_2 \kappa_3 + \kappa_3 \kappa_1 \, ,
\label{Q-0perp}
\end{eqnarray}
where $\sigma$ is a permutation of $(123)$.

The three eigenvalues $\kappa_i$ of the super-stress tensor
$t_{0\perp}$ play for gravitation the same role played by the
principal stresses for electromagnetism. Then, they are the {\it
principal super-stresses}, which are subject to the constraint
\begin{equation}
\kappa_1 + \kappa_2 +\kappa_3 = \xi \, .
\end{equation}

It is worth remarking that in type D the proper super-stress tensor
$t_{0\perp}$ depends on the chosen principal observer. Nevertheless,
the principal super-stresses do not.

\section{Garc\'{\i}a-Parrado radiative gravitational fields}
\label{sec-5}

The super-stress tensor $t_{\perp}$ vanishes only when the Weyl
tensor $W$ vanishes. Thus we can also consider $t_{\perp}$ as a
measure of the gravitational field. Then its flux $Q_{\perp}$
denotes the presence of gravitational radiation. This fact has been
pointed out by Garc\'{\i}a-Parrado \cite{garcia-parrado}, who has
given the following definition.

\begin{definition} \label{def-gapa-g}
{\bf (Intrinsic super-energy radiation, Garc\'{\i}a-Parrado 2008)}
In a vacuum spacetime there exists {\em intrinsic super-energy
radiation} (at a point) if the stress flux tensor $Q_{\perp}$ does
not vanish for any observer.
\end{definition}

For any observer we have $\tr Q_{\perp} = q_{\perp}$. Then,
$q_{\perp}$ vanishes when $Q_{\perp}$ vanishes. Consequently,

\begin{proposition} \label{prop-bel-gp}
Every Bel radiative gravitational field is a Garc\'{\i}a-Parrado
radiative gravitational field.

Every Garc\'{\i}a-Parrado non-radiative gravitational field is a Bel
non-radiative gravitational field, and the observers who do not see
stress flux ($Q_{\perp}=0$) are the observers at rest with respect
to the field.
\end{proposition}

Thus, the definition given by Garc\'{\i}a-Parrado is less
restrictive than Bel's definition, and it allows type $I$ radiative
gravitational fields \cite{garcia-parrado}. Now we analyze them and
consider several significant classes of both radiative and
non-radiative Garc\'{\i}a-Parrado gravitational fields.

\vspace{2mm}

\subsection{Super-energy non radiative gravitational fields}
\label{subsec-5a}

According to the proposition above, the Garc\'{\i}a-Parrado non
radiative fields are type $I$ or type $D$ metrics with a vanishing
proper stress flux tensor $Q_{0 \perp}$, or equivalently, with a
vanishing $\xi_{Q}$. This condition is equivalent to the
Bel-Robinson tensor having real eigenvalues as a consequence of
expressions (\ref{Q-0perp}) and (\ref{ti-pi}). On the other hand
Garc\'{\i}a-Parrado \cite{garcia-parrado} showed that $Q_{0
\perp}=0$ if, and only if, the proper electric and magnetic Weyl
tensors are linearly dependent. Thus we can state.

\vspace{5mm}

\begin{theorem} \label{theo-GP-non}
The Garc\'{\i}a-Parrado non radiative gravitational fields are the
type $I$ or type $D$ metrics which satisfy one of the following
equivalent conditions:

\begin{itemize}
\item[\phantom{iii}(i)] The proper stress flux tensor vanishes, $\; Q_{0 \perp}=0$.
\item[\phantom{ii}(ii)] The proper electric and magnetic Weyl tensors are linearly
dependent, \\$E_0 \otimes H_0 = H_0 \otimes E_0$.
\item[\phantom{i}(iii)] The proper scalar of the stress flux vanishes, $\; \xi_Q
\equiv \sqrt{2 \xi^2 - \frac12 \alpha^2}= 0$.
\item[\phantom{i}(iv)] The Bel-Robinson tensor has real eigenvalues.
\end{itemize}
\end{theorem}

Note that the characterizations (i) and (ii) of the above theorem
make reference to relative quantities, namely the stress flux tensor
and the electric and magnetic parts of the Weyl tensor.
Nevertheless, the conditions (iii) and (iv) are intrinsic in the
Bel-Robinson tensor $T$ and they impose respectively, the vanishing
of an invariant scalar and an algebraic property of $T$.

As pointed out previously \cite{garcia-parrado} all the type $D$
gravitational fields satisfy conditions of theorem
\ref{theo-GP-non}. Now we study the type $I$ metrics which satisfy
them.

Elsewhere \cite{fsWaligned} we have studied the aligned Weyl fields,
i.e. the spacetimes with linearly dependent electric and magnetic
Weyl tensors, and we have shown that they correspond to metrics of
types $IM^+$ or $IM^{\infty}$ in the classification of
McIntosh-Arianrhod \cite{mcintosh}. This means that the scalar
invariant $M$ defined in \ref{M}  is, respectively, a positive real
number or infinity, and they are the type I spacetimes where the
four null Debever directions span a 3-plane \cite{mcintosh, fs-I,
fsWaligned}. Moreover, the (spatial) direction orthogonal to this
3-plane is the Weyl principal direction associated with the Weyl
eigenvalue with the shortest modulus.

We can state this last condition in terms of the principal
super-stresses $\kappa_i$. Indeed, say $\rho_1$ is the shortest Weyl
eigenvalue, then $t_1$ is the shortest Bel-Robinson real eigenvalue
and, from (\ref{t-0perp}), $\kappa_1$ is the largest principal
super-stress. Thus, we have:
\begin{proposition} \label{prop-GP-I-non}
A type I spacetime defines a super-energy non radiative
gravitational field if, and only if, the four null Debever
directions span a 3-plane.

Moreover, the direction orthogonal to this 3-plane is that defined
by the eigenvector associated with the largest principal
super-stress.
\end{proposition}

Finally, we analyze how to distinguish the three types of non
radiative fields in terms of relative super-energy quantities. The results
above imply that the all three cases, $IM^+$, $IM^{\infty}$ and $D$,
are subject to the same constraints for the super-energy scalars:
$$
\xi_q = \xi_Q = 0 \, , \qquad   \xi_T = 3 \xi_t = 3 \xi   \, .
$$
Nevertheless, we can distinguish these three types by using the
principal super-stresses $\kappa_i$. In type $D$ we have $t_2=t_3
\not=0$, $t_1 = 4t_2$, and then $\kappa_2=\kappa_3 = - 2
\kappa_1\not=0$ as a consequence of (\ref{t-0perp}). In type
$IM^{\infty}$ we have $t_1=t_2 \not=0$, $t_3 = 0$, and then
$\kappa_1=\kappa_2 = 0$.

\begin{proposition} \label{prop-GP-I-non-b}
In a super-energy non radiative spacetime the principal
super-stresses satisfy
$$
\kappa_1 \leq 0 \leq \kappa_2 \leq \kappa_3  \, , \qquad |\kappa_1|
\leq |\kappa_2| \, .
$$
Moreover, the spacetime is:

Type $D$ iff $\, \kappa_1 = - \frac12 \kappa_2 < 0 < \kappa_2 =
\kappa_3$.

Type $IM^{\infty}$ iff $\, \kappa_1 = 0 = \kappa_2 < \kappa_3$.

Type $IM^{+}$ otherwise, and then $\kappa_1 < 0 < \kappa_2 <
\kappa_3$.
\end{proposition}

We can quote some examples of non radiative fields:

(i) All the type $D$ vacuum solutions are known \cite{kin, kramer}.

(ii) A result inferred by Szekeres \cite{sze} and confirmed by Brans
\cite{brans} states that no vacuum solutions of type $IM^{\infty}$
exist (see \cite{fs-I1} for a generalization).

(iii) The purely electric or purely magnetic solutions are of type
$IM^{+}$ \cite{fs-EB}. Every static solution has a purely electric
Weyl tensor and, consequently, is of type $IM^{+}$. Both static and
non static purely electric metrics can be found in the Kasner vacuum
solutions \cite{kasner, kramer}. Some restriction are known on the
existence of other $IM^{+}$ vacuum solutions (see \cite{fsWaligned}
and references therein).

\subsection{Super-energy radiative gravitational fields}
\label{subsec-5b}

According to propositions \ref{prop-bel-gp} and \ref{prop-GP-I-non},
the Garc\'{\i}a-Parrado radiative fields are the Bel radiative
fields (types $N$, $III$ and $II$) and the type $I$ metrics with a
non vanishing proper stress flux tensor $Q_{0 \perp}$. According to
theorem \ref{theo-GP-non}, radiative type I fields can be
characterized by the following equivalent conditions: (i) the proper
electric and magnetic Weyl tensors are linearly independent, (ii)
the scalar of the stress flux does not vanish, (iii) the
Bel-Robinson tensor has some complex eigenvalue, (iv) The four null
Debever directions define a frame. Now we consider some relevant
classes of type $I$ radiative fields.

The invariant classification of the Bel-Robinson tensor
\cite{fsBR-1, fsBR-2} leads to three classes with non real
eigenvalues (see appendix \ref{A-Bel-Robinson}): the most regular
one $I_r$, and those with a double or a triple degeneration, which
correspond to types $IM^-$ or $IM^{-6}$, respectively, in the
classification of McIntosh-Arianrhod \cite{mcintosh}. Moreover, they
are the type $I$ spacetimes where the four null Debever directions
define a frame. A frame of vectors is said to be symmetric if they
cannot be distinguished by their mutual metric products
\cite{coll-morales}. In \cite{fsWaligned} we showed that the four
null Debever directions define a symmetric frame for the case
$IM^{-6}$, and a partially symmetric frame for $IM^-$. Thus, we
have:
\begin{proposition} \label{prop-GP-I-rad}
A type I spacetime defines a super-energy radiative gravitational
field if, and only if, the four null Debever vectors define a
frame.

Moreover, this frame is symmetric for type $IM^{-6}$ spacetimes, and
it is partially symmetric (symmetric by pairs) for type $IM^{-}$
spacetimes. Otherwise, the spacetime is of regular type $I_r$.
\end{proposition}

We can label the two degenerate classes in terms of the principal
super-stresses $\kappa_i$. Indeed, from (\ref{t-0perp}), condition
$t_1=t_2=t_3$ implies three equal principal super-stresses,
$\kappa_1=\kappa_2=\kappa_3$, and $t_1\not=t_2=t_3$ implies two
equal principal super-stresses, $\kappa_1\not=\kappa_2=\kappa_3$.

\begin{proposition} \label{prop-GP-I-rad-b}
A super-energy radiative type $I$ spacetime is

Type $IM^{-6}$ iff $\, \kappa_1 = \kappa_2 = \kappa_3$.

Type $IM^{-}$ iff $\, \kappa_1 = \kappa_2 \not= \kappa_3$.

Type $I_r$ otherwise.
\end{proposition}

We can quote some examples of radiative fields:

(i) The Petrov homogeneous vacuum solution \cite{petrov, kramer} is
of type $IM^{-6}$.

(ii) The windmill metric \cite{windmill, kramer} is a family of type
$IM^{-}$ vacuum solutions.

(iii) Vacuum solutions of generic radiative type $I_r$ are quite
abundant. We have, for example, the Taub family of metrics
\cite{taub, kramer} and its counterpart with timelike orbits, or its
related windmill solutions \cite{fs-I1}.

\section{An analysis of type $I$ classes}
\label{sec-6}

The 'degenerate' type $I$ classes defined in \cite{mcintosh} in
terms of the adimensional invariant $M$ have a nice interpretation
in terms of the null Debever directions \cite{mcintosh, fs-I,
fsWaligned}. Moreover, these classes also appear when classifying
the Bel-Robinson tensor as an endomorphism on the nine-dimensional
space of the trace-less symmetric tensors \cite{fsBR-1, fsBR-2}.

In the present paper we have shown how Garcia-Parrado radiative and
non radiative fields can be distinguished in terms of the invariant
$M$. Figure 1(a) presents the complex plane where we can plot the
different values of $M$. Out of the real axes, points correspond to
generic radiative fields $I_r$. For real negative values of $M$ we
have the degenerate radiative fields $IM^{-}$ and $IM^{-6}$. The
real non negative values of $M$ correspond to non radiative fields:
type $D$ when $M=0$, type $IM^{+}$ when $M>0$ and type $IM^{\infty}$
when $M$ is not defined.

The study of the different classes using the Bel-Robinson tensor
\cite{fsBR-1, fsBR-2} implies analyzing the Bel-Robinson real
eigenvalues $t_i$. And an approach using the super-energy relative
magnituds can be built with the principal super-stresses $\kappa_i$.

\subsection{A diagram approach using Bel-Robinson real eigenvalues}

\begin{figure}
\centerline{
\parbox[c]{0.78\textwidth}{\includegraphics[width=0.88\textwidth]{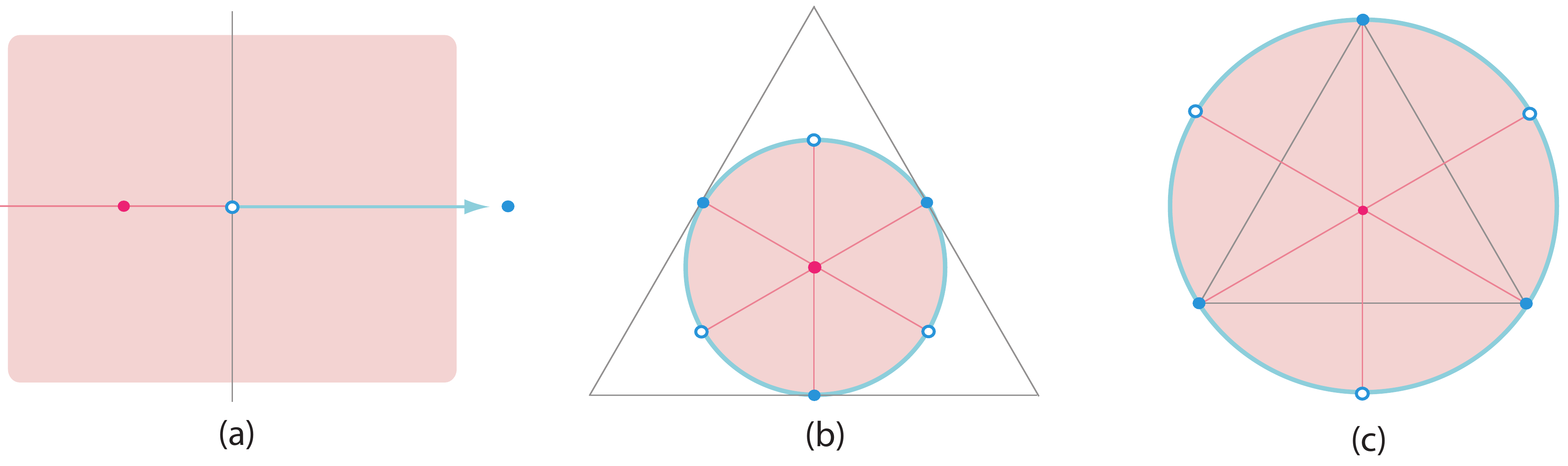}}}
\caption{In these diagrams reddish points represent radiative
metrics: red solid dot type $IM^{-6}$, red lines type $IM^{-}$, and
type $I_r$ otherwise. And bluish points represent non radiative
metrics: blue solid dot type $IM^{\infty}$, blue empty dot type $D$,
and blue lines type $IM^{+}$. (a) Plane where the complex invariant
$M$ is plotted. (b) Plane $\bar{t}_1 + \bar{t}_2 + \bar{t}_3 = 1$ in
the parameter space $\{\bar{t}_1, \bar{t}_2, \bar{t}_3\}$. (c) Plane
$\bar{\kappa}_1 + \bar{\kappa}_2 + \bar{\kappa}_3 = 1$ in the
parameter space $\{\bar{\kappa}_1, \bar{\kappa}_2,
\bar{\kappa}_3\}$. In diagrams (b) an (c) the black triangle
represents points on the corresponding coordinate planes.
\label{Fig-TypeIM}}
\end{figure}

Constraints on the real eigenvalues $t_i$:
$$
t_1 + t_2 + t_3 = 4 \xi \, , \quad \quad  t_1^2 + t_2^2 + t_3^2 = 4
\chi \leq 8 \xi^2 \, ,
$$
Then, the adimensional parameters $ \displaystyle \bar{t}_i =
\frac{t_i}{4 \xi}$ satisfy:
$$
\bar{t}_1 + \bar{t}_2 + \bar{t}_3 = 1 \, , \quad \quad \bar{t}_1^2 +
\bar{t}_2^2 + \bar{t}_3^2 \leq \frac12 \, ,
$$
conditions which represent the points on a plane $\Pi$, and in a
sphere $S$, respectively, in the parameter space $\{\bar{t}_1,
\bar{t}_2, \bar{t}_3\}$. Every type $I$ metric corresponds to a
point on the circle surrounded by the intersection circumference $C
= \Pi \cap S$. This $C$ is the incircle of triangle $T$ on $\Pi$
defined by coordinate planes $ \bar{t}_1 \bar{t}_2   \bar{t}_3 = 0$
(see Figure \ref{Fig-TypeIM}(b)). The non radiative fields  belong
to the circumference $C$, and the radiative fields are in its
interior. The degenerate both non radiative and radiative fields are
located on the medians of the triangle $T$: type $IM^{\infty}$ on
the base points, type $D$ on the three opposing points in $C$, type
$IM^{-6}$ on the barycenter and $IM^{-}$ in any other point on the
medians.

\subsection{A diagram approach using principal super-stress scalars}

Constraints on the principal super-stress scalars $\kappa_i$:
$$
\kappa_1 + \kappa_2 +\kappa_3 = \xi \, , \quad \quad \kappa_1^2 +
\kappa_2^2 + \kappa_3^2 =  \xi^2 - 2 \kappa^2 \leq \xi^2 \, ,
$$
Then, the adimensional parameters $ \displaystyle \bar{\kappa}_i =
\frac{\kappa_i}{\xi}$ satisfy:
$$
\bar{\kappa}_1 + \bar{\kappa}_2 + \bar{\kappa}_3 = 1 \, , \quad
\quad  \bar{\kappa}_1^2 + \bar{\kappa}_2^2 + \bar{\kappa}_3^2 \leq 1
\, ,
$$
conditions which represent the points on a plane $\Pi$, and in a
sphere $S$, respectively, in the parameter space $\{\bar{\kappa}_1,
\bar{\kappa}_2, \bar{\kappa}_3\}$. Every type $I$ metric corresponds
to a point on the circle surrounded by the intersection
circumference $C = \Pi \cap S$. This $C$ is the circumcircle of
triangle $T$ on $\Pi$ defined by coordinate planes $\bar{\kappa}_1
\bar{\kappa}_2 \bar{\kappa}_3 = 0$  (see Figure
\ref{Fig-TypeIM}(c)). The non radiative fields belong to the
circumference $C$, and the radiative fields are in its interior.
Both non radiative and radiative degenerate fields are located on
the medians of triangle $T$: type $IM^{\infty}$ on the vertex
points, type $D$ on the three opposing points in $C$, type $IM^{-6}$
on the barycenter and $IM^{-}$ in any other point on the medians.

We can see in the diagram that non radiative types $D$ and
$IM^{\infty}$ can be considered as the limit of the 'generic' non
radiative type $IM^{+}$, but can also be the limit of radiative
cases, in particular of the 'degenerate' radiative type $IM^{-}$.
Type $IM^{-6}$ is the farthest from the non radiative types, that is
to say, it could be considered the most radiative type $I$ case.
This statement, based on geometric considerations, can also be
supported by an analytical approach. Indeed, if we define the {\it
adimensional radiation parameter} $\bar{\kappa}^2 =
\frac{\kappa^2}{\xi^2}$, we have: $\bar{\kappa}^2$ vanishes for non
radiative types and it reaches the maximum value for the radiative
type $IM^{-6}$. The study of this or other similar radiation
parameters for known vacuum solutions would be an interesting task
which we will undertake elsewhere.

\ack We thank A Garc\'{\i}a-Parrado and J A Morales-Lladosa for
their comments. We are grateful to the referees for their valuable
suggestions. This work has been supported by the Spanish ministries
of ``Ciencia e Innovaci\'on" and ``Econom\'{\i}a y Competitividad",
MICINN-FEDER projects FIS2009-07705 and FIS2012-33582.

\appendix

\section{Algebraic classification of the Bel-Robinson tensor}
\label{A-Bel-Robinson}

The classification of the Bel-Robinson tensor $T$ as an endomorphism
\cite{fsBR-1, fsBR-2} leads to nine classes: the Petrov-Bel types
$O$, $N$, $III$ and $II$ and five subclasses of type $I$ metrics:
types $I_r$, $IM^-$, $IM^{-6}$, $IM^+$ and $IM^{\infty}$. These
'degenerate' type I metrics may be characterized in terms of the
adimensional Weyl scalar invariant \cite{mcintosh, fs-I}:
\begin{equation} \label{M}
M = \frac{a^3}{b^2}-6
\end{equation}
where $a$ and $b$ are the quadratic and the cubic Weyl complex
scalar invariants. Now we summarize for every class the degeneration
of the nine Bel-Robinson eigenvalues $\{ t_1, t_2, t_3, \tau_1,
\tau_2, \tau_3, \bar{\tau}_1 , \overline{\tau}_2, \bar{\tau}_3 \}$:

\begin{description}
\item[]
{\em Type $I_r$.} This is the more regular case. The scalar $M$ is
not real, and $T$ has nine different eigenvalues, three real ones
and three pairs of complex conjugate.
\item[]
{\em Type $IM^-$}. In this case $M$ is a negative real number
different from $-6$, and $T$ has six different eigenvalues, two real
ones, a simple and a double, and two double and two simple complex
conjugate eigenvalues: $t_1 = t_2$ and $\tau_1 = \tau_2 \not=0$.
\item[]
{\em Type $IM^{-6}$}. In this case $M$ is the real number $-6$, and
$T$ has three triple eigenvalues, one real and a pair of complex
conjugate:  $t_1 = t_2 =t_3$ and $\tau_1 = \tau_2 = \tau_3$.
\item[]
{\em Type $IM^+$}. In this case $M$ is a positive real number, and
$T$ has six different real eigenvalues, three simple ones and three
double ones: $\tau_i = \bar{\tau_i}$.
\item[]
{\em Type $IM^{\infty}$}. In this case $M$ is infinity, and $T$ has
three different real eigenvalues with multiplicities 2, 2, 5: $t_1 =
t_2 \not=0$, $\tau_3 = \bar{\tau}_3 = -t_1$ and $t_3 = \tau_1 =
\tau_2=0$.
\item[]
{\em Type $D$}. In this case $M$ vanishes, and the eigenvalues of
$T$ are restricted by $t_2 = t_3 = \tau_1 \not=0$, $t_1 = 4 t_2$ and
$\tau_2 = \tau_3 = -2t_2$.
\item[]
{\em Type $II$}. This case only differs of type $D$ in the minimal
polynomial, with eigenvalues of $T$ restricted by $t_2 = t_3 =
\tau_1 \not=0$, $t_1 = 4 t_2$ and $\tau_2 = \tau_3 = -2t_2$.
\item[]
{\em Type $III$}. In this case all the Bel-Robinson eigenvalues
vanish.
\item[]
{\em Type $N$}. This case only differs of type $III$ in the minimal
polynomial, and all the eigenvalues vanish.
\end{description}

\section{Some constraints on the super-energy scalars}
\label{B-constraints}

From expressions (\ref{relative}) and (\ref{T-Q-t-q-tau}) we easily
obtain the following relations between the associated quadratic
scalars:
\begin{equation}
\begin{array}{rcl}
(T,T) & = & \tau^2 - 4\, |q_{\perp}|^2 + 6 \, |t_{\perp}|^2  - 4\,
|Q_{\perp}|^2  + |T_{\perp}|^2  \, , \cr (Q,Q) & = & -\tau^2 + 3 \,
|q_{\perp}|^2  - 3\, |t_{\perp}|^2  + |Q_{\perp}|^2  \, , \cr (t,t)
& = & \tau^2 - 2\, |q_{\perp}|^2 + |t_{\perp}|^2  \, , \cr (q,q) & =
& - \tau^2 + |q_{\perp}|^2 \, .
\end{array}
\end{equation}

The scalars $(T,T)$ and $(Q,Q)$ are invariants as stated in
(\ref{sei}). Moreover, from the Bergqvist and Lankinen conditions
\cite{bergqvist-lan} we also obtain:
\begin{equation} \label{bergqvist}
3\,(t,t) + 4\, (q,q) = \frac12 \alpha^2
\end{equation}
Consequently, the super-energy scalars are subject to the\\
 {\it Quadratic scalar constraints}
\begin{equation} \label{qsc}
\begin{array}{rcl}
4 \alpha^2   & = & \tau^2 - 4\, |q_{\perp}|^2 + 6 \, |t_{\perp}|^2
- 4\, |Q_{\perp}|^2  + |T_{\perp}|^2  \, , \cr  \alpha^2  & = &
\tau^2 - 3 \, |q_{\perp}|^2  + 3\, |t_{\perp}|^2  - |Q_{\perp}|^2 \,
, \cr  \frac12 \alpha^2 & = & - \tau^2 - 2\, |q_{\perp}|^2 + 3\,
|t_{\perp}|^2\, .
\end{array}
\end{equation}

On the other hand, from the Bergqvist and Lankinen conditions
\cite{bergqvist-lan} we obtain the following
\\
 {\it Quadratic vectorial constraints} \label{qvc}
\begin{equation}
\begin{array}{rcl}
Q_\perp (t_\perp) - \tau \, q_\perp & = & 0    \, , \cr
T_\perp (Q_\perp)  - 3 Q_\perp
(t_\perp) + 3 t_\perp (q_\perp)  - \tau \, q_\perp &  = & 0
\, .
\end{array}
\end{equation}
 {\it Quadratic 2-order tensorial constraints}
\begin{equation}
\hspace{-10mm}
\begin{array}{rcl}
T_\perp(t_\perp) + Q_\perp  \! \cdot^{\hspace{-1.3mm} 2} Q_\perp + 2
Q_\perp   (q_\perp) - 3 \tau \, t_\perp + 2 \, t_\perp \! \cdot
t_\perp - 3 q_\perp \otimes q_\perp  & = & 0  \, , \cr Q_\perp  \!
\cdot^{\hspace{-1.3mm} 2} Q_\perp -  q_\perp \otimes q_\perp -
(\tau^2 -  |t_{\perp}|^2) \gamma & = & 0
 \, , \cr  T_{\perp}   \! \cdot^{\hspace{-1.3mm} 3} T_\perp - 3 Q_\perp  \!

\cdot^{\hspace{-1.3mm} 2} Q_\perp + 3 t_\perp \! \cdot t_\perp -
q_\perp \otimes q_\perp - \alpha^2 \, \gamma& = & 0 \, ,
\end{array}
\end{equation}
A dot $\cdot$ denotes the contraction of adjacent indices.
Similarly, $ \cdot^{\hspace{-1.3mm} 2}$ and $\cdot^{\hspace{-1.3mm}
3}$ denote, respectively, a double and a triple contraction.

\section{}
\label{C-propositions}
%
{\em Proof of proposition \ref{prop-tt}}. This Proposition states
that the following relation holds:
\begin{equation} \label{tt-chi}
(t,t) \geq  \chi \geq  \frac{1}{2} \alpha^2 \geq 0 ,
\end{equation}
and moreover, it specifies when each of the three involved inequalities becomes
strict or is an equality depending on the different Petrov-Bel
types.

In types $N$ and $III$ the Bel-Robinson eigenvalues vanish. Then,
expressions (\ref{xi-chi}) imply $\chi = \frac12 \alpha = 0$.

In types $D$ and $II$ the real Bel-Robinson eigenvalues satisfy $t_2
= t_3 = \tau_1 \not=0$, $t_1 = 4 t_2$. Then, expressions
(\ref{xi-chi}) imply $\chi = \frac92 t_2^2 = \frac12 \alpha \not=
0$.

In type $I$ we obtain, from the Bel-Robinson canonical form (see
\cite{fs-RS}),
$$
16\, \chi^2\! = \! \Big(\!\sum_{i=1}^{3}\! t_i^2 \Big)^2 \! \! =
\sum_{i=1}^{3}\! t_i^4 +2 \sum_{i < j}\! t_i^2 t_j^2 = \!
\sum_{i=1}^{3}\! t_i^4 + 2 \sum_{k=1}^{3}\! | \tau_k|^4 \geq \!
\sum_{i=1}^{3}[ t_i^4 +  \tau_i^4 +  \bar{\tau}_i^4] =\tr T^4 = 4
\alpha^4 \, ,
$$
where the last equality has been proved in \cite{fsBR-1}, $\tr T^4$
denoting the trace of the fourth power of $T$ as an endomorphism.

From the definition (\ref{T-Q-t-q-tau}) of $t$, the first inequality
in (\ref{tt-chi}) writes $\ (t,t) = T^2(u,u,u,u) \geq \chi \, $.

In type $N$, $T^2 = 0$, and then $T^2(u,u,u,u)=0=\chi$.

In type $III$, the canonical form (see \cite{fs-RS}) implies
$T^2(u,u,u,u) = (l,x)^2 > 0 = \chi$.

In type $II$, an arbitrary observer $u$ in terms of the Bel-Robinson
canonical frame takes the expression $u = \lambda(e^{\phi} \ell +
e^{-\phi} k) + \mu (e^{\ci \sigma} m + e^{-\ci \sigma} \bar{m})$,
$2(\lambda^2 - \mu^2) = 1$. Then, from the Bel-Robinson canonical
form (see \cite{fs-RS}) we obtain
$$
T^2(u,u,u,u)  = \chi \left[1 + 4 \mu^2 + 2 \mu^4 \sin^2 2 \sigma + 8
\left(\frac13 \lambda^2 e^{-2 \phi} \! - \mu^2 \cos 2 \sigma
\right)^2\right] > \chi \, .
$$

Finally, we study types $I$ and $D$.  In \cite{fsBR-1} we have
introduced a second order super-energy tensor $T_{(2)}$ associated
with the traceless part $W_{(2)}$ of the square $W^2$ of the Weyl
tensor $W$. That is, $T_{(2)}$ is defined as (\ref{BR-1}) by
changing $W$ by $W_{(2)}$. It follows that $T_{(2)}$ has the same
properties as $T$ \cite{fsBR-1}. Then, we can apply to it the last
inequality in expression (\ref{sei}) of theorem \ref{theo-sei}: if
$e_0$ is a principal observer, for any observer $u$ we have:
\begin{equation} \label{T(2)}
T_{(2)}(u,u,u,u) \geq  T_{(2)}(e_0,e_0,e_0,e_0)  \, .
\end{equation}
But, for any observer $u$, $T_{(2)}(u,u,u,u) = T^2(u,u,u,u) -\frac13
\alpha^2$. Thus, (\ref{T(2)}) holds by substituting $T_{(2)}$ by
$T^2$, and we obtain:
\begin{equation} \label{T2}
T^2(u,u,u,u) \geq  T^2(e_0,e_0,e_0,e_0)  = \frac14
\!\sum_{i=1}^{3}\! t_i^2  = \chi \, .
\end{equation}

\noindent
{\em Proof of proposition \ref{prop-qq}}. This proposition states that the following relation holds:
\begin{equation} \label{qq-xi}
-(q,q) \geq  \xi^2 \geq  \frac{1}{4} \alpha^2 \geq 0 ,
\end{equation}
and moreover, it specifies when each of the three involved
inequalities becomes strict or is an equality depending on the
different Petrov-Bel types.

From (\ref{xi-chi}), (\ref{bergqvist}) and (\ref{tt-chi}) we obtain
$$
-4(q,q) = 3(t,t) - \frac12 \alpha^2 \geq 3 \chi -\frac12 \alpha = 4
\xi^2 \geq \alpha^2 \, .
$$
Moreover, every inequality becomes strict (or an equality) when the
corresponding inequality in (\ref{tt-chi}) becomes strict (or an
equality).

\end{document}